\documentclass[letterpaper,10pt,conference,dvipsnames]{ieeeconf}
\IEEEoverridecommandlockouts
\usepackage{amsmath,amssymb,amsfonts}
\usepackage{bm}
\usepackage{cite}
\usepackage{transparent} 
\usepackage{graphicx}
\usepackage{booktabs}
\usepackage{dsfont}
\usepackage{tikz}
\usepackage{svg}
\usepackage{flushend}
\usetikzlibrary{shapes,arrows,positioning,calc}

\usepackage[table]{xcolor}
\usepackage{booktabs,tabularx}
\definecolor{simTableHead}{RGB}{227,235,242}
\definecolor{simTableStripe}{RGB}{245,248,251}

\usepackage{hyperref}
\hypersetup{
    colorlinks=true,
    linkcolor=MidnightBlue,
    citecolor=PineGreen,
    urlcolor=Tan,
    pdftitle={Predictive Neuromorphic Control of A Scanning Mirror System},
    pdfauthor={Changrui Liu}
}

\AtBeginDocument{
  \setlength{\textfloatsep}{10pt}
  \setlength{\intextsep}{10pt}
}

\newtheorem{remark}{Remark}

\newcommand{\sminus}{\hspace{-0.06cm}-\hspace{-0.06cm}}
\newcommand{\seq}{\hspace{-0.08cm}=\hspace{-0.08cm}}

\newcommand{\sleq}{\hspace{-0.08cm}\leq\hspace{-0.08cm}}

\newcommand{\pecmpc}{\mathrm{P}_{\mathrm{EC-MPC}}}

\newcommand{\mvec}{\texttt{vec}}
\newcommand{\mdiag}{\texttt{diag}}

\newcommand{\dt}{\mathrm{d}t}

\newcommand{\dsone}{\mathds{1}}

\newcommand{\bR}{\mathbb{R}}
\newcommand{\bN}{\mathbb{N}}
\newcommand{\bI}{\mathbb{I}}
\newcommand{\cP}{\mathcal{P}}
\newcommand{\cA}{\mathcal{A}}
\newcommand{\cX}{\mathcal{X}}

\newcommand{\Mm}{\mathcal{M}_{\mathrm{m}}}
\newcommand{\Mc}{\mathcal{M}_{\mathrm{c}}}

\newcommand{\myF}{F_{\Mm,\Mc,\mathcal{E}_{\mathrm{act}}}}

\newcommand{\rmL}{\mathrm{L}}
\newcommand{\rmM}{\mathrm{M}}
\newcommand{\rmR}{\mathrm{R}}

\newcommand{\Em}{E_{\mathrm{M}}}
\newcommand{\El}{E_{\mathrm{L}}}
\newcommand{\Er}{E_{\mathrm{R}}}
\newcommand{\tm}[1]{t_{\mathrm{M},#1}}

\newcommand{\htm}[1]{\hat{t}_{\mathrm{M},#1}}

\newcommand{\hA}{\hat{\alpha}}
\newcommand{\hbA}{\hat{A}}

\newcommand{\gsm}{g^-_{\mathrm{s}}}

\newcommand{\gus}{g^+_{\mathrm{us}}}

\newcommand{\gsmin}{g^-_{\mathrm{s},\min}}
\newcommand{\gsmax}{g^-_{\mathrm{s},\max}}
\newcommand{\gusmin}{g^+_{\mathrm{us},\min}}
\newcommand{\gusmax}{g^+_{\mathrm{us},\max}}

\newcommand{\gsme}[1]{g^-_{\mathrm{s},(#1)}}
\newcommand{\guse}[1]{g^+_{\mathrm{us},(#1)}}
\newcommand{\Ie}[1]{I_{0,(#1)}}
\newcommand{\thetae}[1]{\theta_{(#1)}}
\newcommand{\tthetae}[1]{\tilde{\theta}^\star_{(#1)}}

\newcommand{\tf}{\tau_{\mathrm{f}}}
\newcommand{\ts}{\tau_{\mathrm{s}}}
\newcommand{\tus}{\tau_{\mathrm{us}}}

\newcommand{\vth}{\nu_{\mathrm{th}}}

\newcommand{\vf}[1]{\nu_{\mathrm{f},#1}}
\newcommand{\vs}[1]{\nu_{\mathrm{s},#1}}
\newcommand{\vus}[1]{\nu_{\mathrm{us},#1}}
\newcommand{\dvf}[1]{\dot{\nu}_{\mathrm{f},#1}}
\newcommand{\dvs}[1]{\dot{\nu}_{\mathrm{s},#1}}
\newcommand{\dvus}[1]{\dot{\nu}_{\mathrm{us},#1}}

\newcommand{\isyn}[2]{I_{\mathrm{syn},#1#2}}
\newcommand{\iext}[1]{I_{\mathrm{ext},#1}}

\title{\LARGE\bfseries Predictive Neuromorphic Control of a Scanning Mirror System: An Event-Centric Framework}

\author{%
Changrui Liu$^{1,2}$,
Robin Wroblowski$^{2}$,
Anil Alan$^{3}$,
Shengling Shi$^{1}$,
Rodolphe Sepulchre$^{2,4}$
Bart De Schutter$^{1}$
\thanks{*This paper is part of a project that has received funding from the European Research Council (ERC) under the European Union’s Horizon 2020 research and innovation programme (Grant agreement No. 101018826 - CLariNet and No. 101054323 - SpikyControl).}%
\thanks{This work was completed during Changrui Liu's research visit to the STADIUS Center at KU Leuven.}
\thanks{$^{1}$Delft Center for Systems and Control, Delft University of Technology, Mekelweg 2, Delft, The Netherlands. E-mails: \texttt{\{c.liu-14, shengling.shi, b.deschutter\}@tudelft.nl} (Changrui Liu, Shengling Shi, Bart De Schutter)}%
\thanks{$^{2}$STADIUS Center for Dynamical Systems, Signal Processing, and Data Analytics, Department of Electrical Engineering (ESAT), KU Leuven, KasteelPark Arenberg 10, Leuven, Belgium. E-mails: \texttt{\{robin.wroblowski, rodolphe.sepulchre\}@kuleuven.be} (Robin Wroblowski, Rodolphe Sepulchre)}%
\thanks{$^{3}$Control Systems Group, Department of Electrical Engineering, Eindhoven University of Technology, Groene Loper 19, Eindhoven, The Netherlands. E-mail: \texttt{a.alan@tue.nl} (Anil Alan)}%
\thanks{$^{4}$Department of Engineering, University of Cambridge, Trumpington Street, Cambridge CB2 1PZ, United Kingdom.}
}

\begin{document}
\bstctlcite{BSTcontrol}
\maketitle
\thispagestyle{empty}
\pagestyle{empty}

\begin{abstract}
Organizing control around events can reduce sensing and actuation demands while directly linking control decisions to the occurrence and timing of desired system behaviors. This paper proposes a novel event-centric model predictive control (MPC) framework that unifies event sensing, event prediction, and event regulation. Events define the feedback information, prediction horizon, and event-timing objectives, placing them at the center of predictive control design. The framework is developed for a scanning mirror system equipped with a baseline neuromorphic controller composed of spiking neurons that supports sparse actuation. The control objective is to achieve prescribed zero-crossing times and a desired scanning range while minimizing the mechanical energy transferred between the actuator and the mirror. A sampling-based algorithm is presented to obtain suboptimal solutions of the event-centric MPC optimization problem without requiring an explicit event-to-event prediction model. By grounding predictive optimization in events, the proposed framework opens a new direction for MPC design and its integration with neuromorphic control.
\end{abstract}

\section{Introduction}
\label{sec:intro}
Neuromorphic control exploits the dynamics of excitable neuronal circuits to regulate behavior through sensing and actuation events~\cite{ribar2021neuromorphic}. It is particularly relevant to rhythmic systems, where control relies on coordinating recurring events, with applications including pendulum control~\cite{schmetterling2024neuromorphic} and (soft) robotic locomotion~\cite{arbelaiz2025spiking}. Scanning mirrors provide another application of rhythmic control, generating repetitive motion to steer optical beams for confocal microscopy~\cite{ji2008advances}, laser micro-machining~\cite{de2013polygon}, and light detection and ranging (LiDAR)~\cite{kimoto2014development}. To achieve repeatable beam steering, the mirror should sustain oscillations at a desired scanning amplitude and with precisely controlled timing. Existing control approaches for rotational reluctance scanning mirrors combine zero-crossing synchronization with amplitude regulation based on linearized models~\cite{pechgraber2023resonant}. However, these approaches do not address the joint regulation of prescribed future zero-crossing times and the scanning range while minimizing the mechanical energy transferred between the actuator and the mirror under sparse feedback. The rhythmic nature of these systems makes neuromorphic control a promising approach, but a key challenge remains: systematically tuning the neuronal parameters to satisfy multiple control objectives.
\begin{figure}[t]
    \vspace{0.6\baselineskip} 
    \centering
    \resizebox{\columnwidth}{!}{%

\tikzset{every picture/.style={line width=0.75pt}} 

\begin{tikzpicture}[x=0.75pt,y=0.75pt,yscale=-1,xscale=1]

\draw  [fill={rgb, 255:red, 155; green, 155; blue, 155 }  ,fill opacity=0.1 ][dash pattern={on 4.5pt off 4.5pt}] (51,35) -- (308,35) -- (308,197) -- (51,197) -- cycle ;
\draw  [color={rgb, 255:red, 184; green, 233; blue, 134 }  ,draw opacity=1 ][fill={rgb, 255:red, 184; green, 233; blue, 134 }  ,fill opacity=0.2 ][line width=1.5]  (200,103) .. controls (200,95.27) and (206.27,89) .. (214,89) .. controls (221.73,89) and (228,95.27) .. (228,103) .. controls (228,110.73) and (221.73,117) .. (214,117) .. controls (206.27,117) and (200,110.73) .. (200,103) -- cycle ;
\draw  [color={rgb, 255:red, 80; green, 227; blue, 194 }  ,draw opacity=1 ][fill={rgb, 255:red, 80; green, 227; blue, 194 }  ,fill opacity=0.2 ][line width=1.5]  (262,103) .. controls (262,95.27) and (268.27,89) .. (276,89) .. controls (283.73,89) and (290,95.27) .. (290,103) .. controls (290,110.73) and (283.73,117) .. (276,117) .. controls (268.27,117) and (262,110.73) .. (262,103) -- cycle ;
\draw [color={rgb, 255:red, 245; green, 166; blue, 35 }  ,draw opacity=1 ]   (224,95) .. controls (233.2,87.4) and (252.2,86.4) .. (263.2,95.4) ;
\draw [shift={(263.2,95.4)}, rotate = 39.29] [color={rgb, 255:red, 245; green, 166; blue, 35 }  ,draw opacity=1 ][fill={rgb, 255:red, 245; green, 166; blue, 35 }  ,fill opacity=1 ][line width=0.75]      (0, 0) circle [x radius= 2.01, y radius= 2.01]   ;
\draw [color={rgb, 255:red, 245; green, 166; blue, 35 }  ,draw opacity=1 ]   (225,111) .. controls (237.2,118.4) and (253.2,119.4) .. (263.2,110.4) ;
\draw [shift={(225,111)}, rotate = 31.24] [color={rgb, 255:red, 245; green, 166; blue, 35 }  ,draw opacity=1 ][fill={rgb, 255:red, 245; green, 166; blue, 35 }  ,fill opacity=1 ][line width=0.75]      (0, 0) circle [x radius= 2.01, y radius= 2.01]   ;
\draw  [fill={rgb, 255:red, 248; green, 231; blue, 28 }  ,fill opacity=0.15 ] (192,77.45) .. controls (192,72.78) and (195.78,69) .. (200.45,69) -- (291.55,69) .. controls (296.22,69) and (300,72.78) .. (300,77.45) -- (300,115.83) .. controls (300,120.5) and (296.22,124.29) .. (291.55,124.29) -- (200.45,124.29) .. controls (195.78,124.29) and (192,120.5) .. (192,115.83) -- cycle ;
\draw (432,143) node  {\includegraphics[width=90pt,height=54pt]{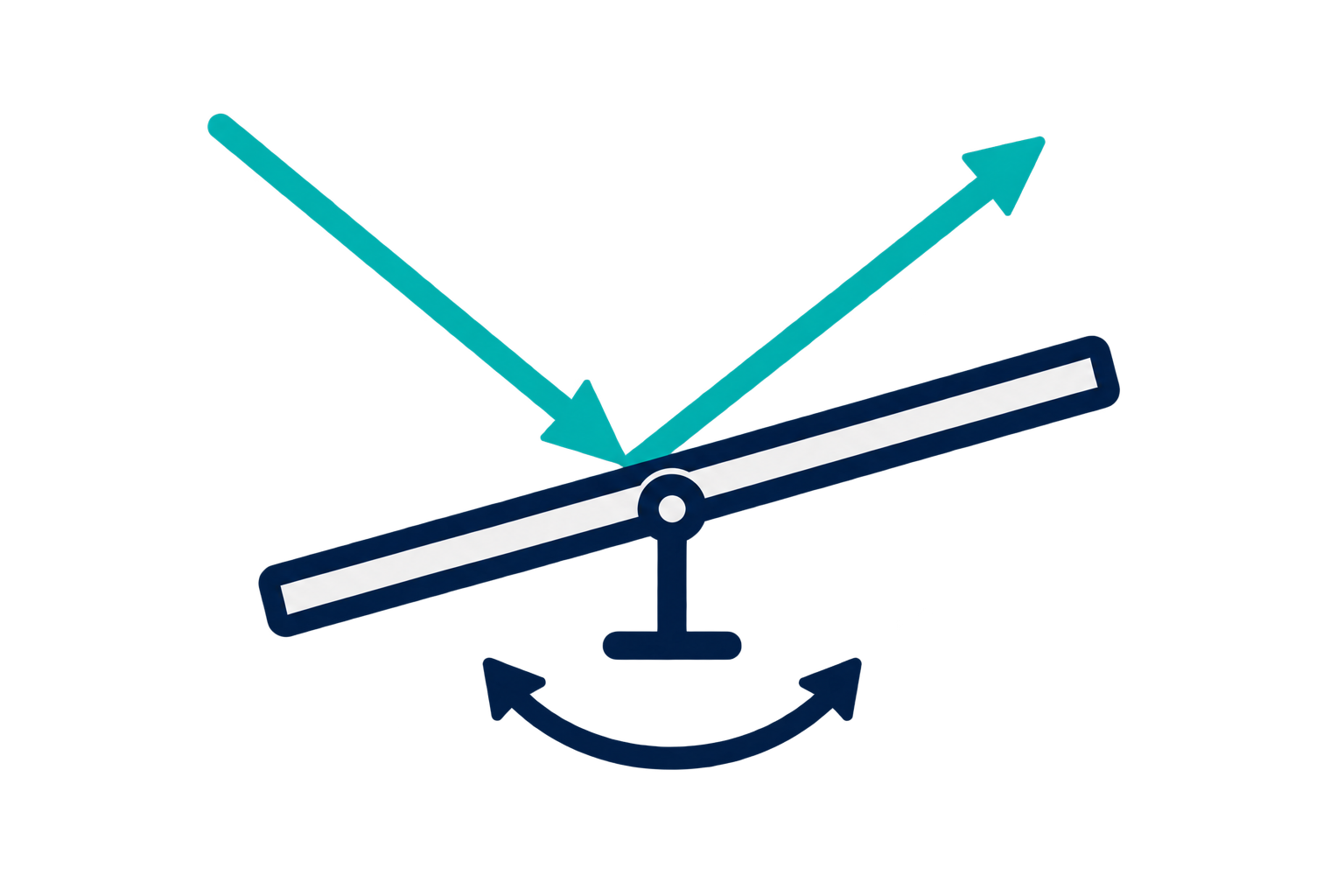}};
\draw  [fill={rgb, 255:red, 245; green, 166; blue, 35 }  ,fill opacity=0.15 ] (382,96) .. controls (382,90.63) and (386.35,86.29) .. (391.71,86.29) -- (468.29,86.29) .. controls (473.65,86.29) and (478,90.63) .. (478,96) -- (478,166.29) .. controls (478,171.65) and (473.65,176) .. (468.29,176) -- (391.71,176) .. controls (386.35,176) and (382,171.65) .. (382,166.29) -- cycle ;
\draw  [fill={rgb, 255:red, 139; green, 87; blue, 42 }  ,fill opacity=0.15 ] (192,149) .. controls (192,145.29) and (195.01,142.29) .. (198.71,142.29) -- (293.29,142.29) .. controls (296.99,142.29) and (300,145.29) .. (300,149) -- (300,184.29) .. controls (300,187.99) and (296.99,191) .. (293.29,191) -- (198.71,191) .. controls (195.01,191) and (192,187.99) .. (192,184.29) -- cycle ;
\draw (247.5,172.5) node [rotate=-90.13] {\includegraphics[width=53.25pt,height=42.75pt]{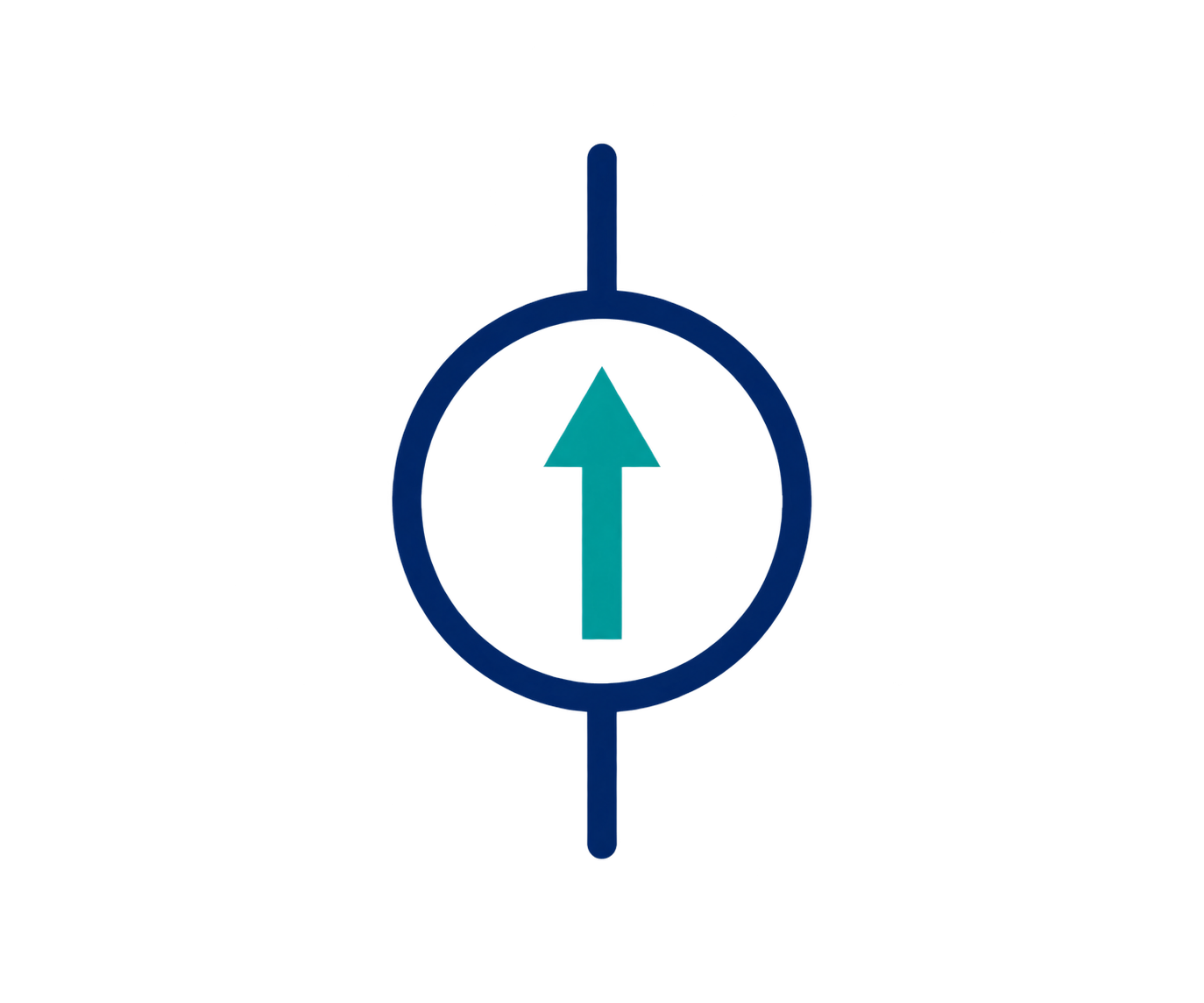}};
\draw [color={rgb, 255:red, 208; green, 2; blue, 27 }  ,draw opacity=1 ]   (244,124.67) -- (244,138.67) ;
\draw [shift={(244,141.67)}, rotate = 270] [fill={rgb, 255:red, 208; green, 2; blue, 27 }  ,fill opacity=1 ][line width=0.08]  [draw opacity=0] (5.36,-2.57) -- (0,0) -- (5.36,2.57) -- cycle    ;
\draw  [fill={rgb, 255:red, 126; green, 211; blue, 33 }  ,fill opacity=0.15 ] (57,98) .. controls (57,92.63) and (61.35,88.29) .. (66.71,88.29) -- (161.29,88.29) .. controls (166.65,88.29) and (171,92.63) .. (171,98) -- (171,165.71) .. controls (171,171.08) and (166.65,175.43) .. (161.29,175.43) -- (66.71,175.43) .. controls (61.35,175.43) and (57,171.08) .. (57,165.71) -- cycle ;
\draw (114.21,137.71) node  {\includegraphics[width=73.82pt,height=44.57pt]{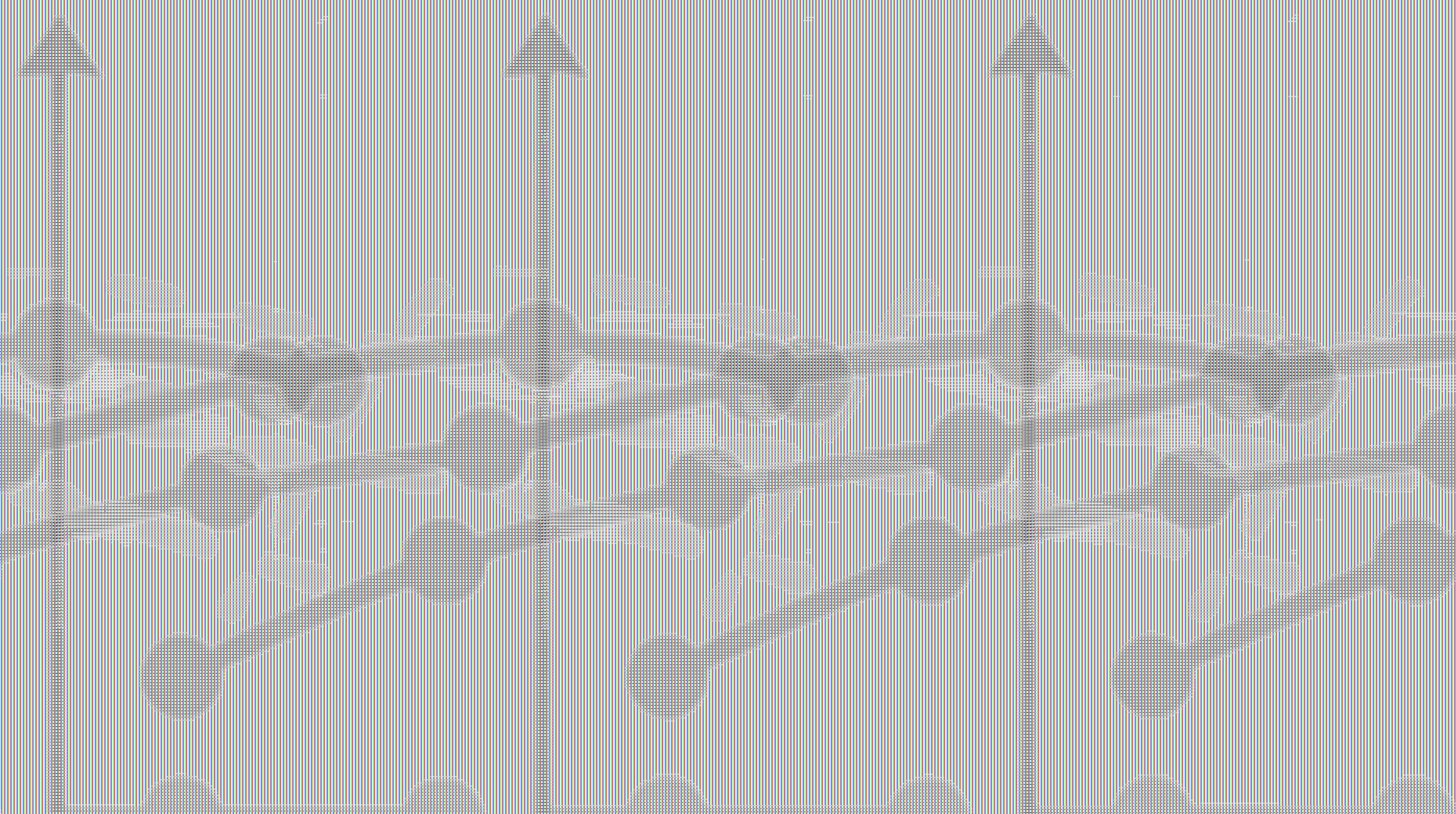}};
\draw    (171,110) -- (189,110) ;
\draw [shift={(192,110)}, rotate = 180] [fill={rgb, 255:red, 0; green, 0; blue, 0 }  ][line width=0.08]  [draw opacity=0] (5.36,-2.57) -- (0,0) -- (5.36,2.57) -- cycle    ;
\draw    (171,100) -- (176,100) ;
\draw    (176,85) -- (189,85) ;
\draw [shift={(192,85)}, rotate = 180] [fill={rgb, 255:red, 0; green, 0; blue, 0 }  ][line width=0.08]  [draw opacity=0] (5.36,-2.57) -- (0,0) -- (5.36,2.57) -- cycle    ;
\draw    (176,85) -- (176,100) ;
\draw    (171,165) -- (189,165) ;
\draw [shift={(192,165)}, rotate = 180] [fill={rgb, 255:red, 0; green, 0; blue, 0 }  ][line width=0.08]  [draw opacity=0] (5.36,-2.57) -- (0,0) -- (5.36,2.57) -- cycle    ;
\draw    (300,165) -- (321,165) ;
\draw    (321,129) -- (321,165) ;
\draw    (321,129) -- (379,129) ;
\draw [shift={(382,129)}, rotate = 180] [fill={rgb, 255:red, 0; green, 0; blue, 0 }  ][line width=0.08]  [draw opacity=0] (5.36,-2.57) -- (0,0) -- (5.36,2.57) -- cycle    ;
\draw (352,139) node  {\includegraphics[width=39pt,height=15pt]{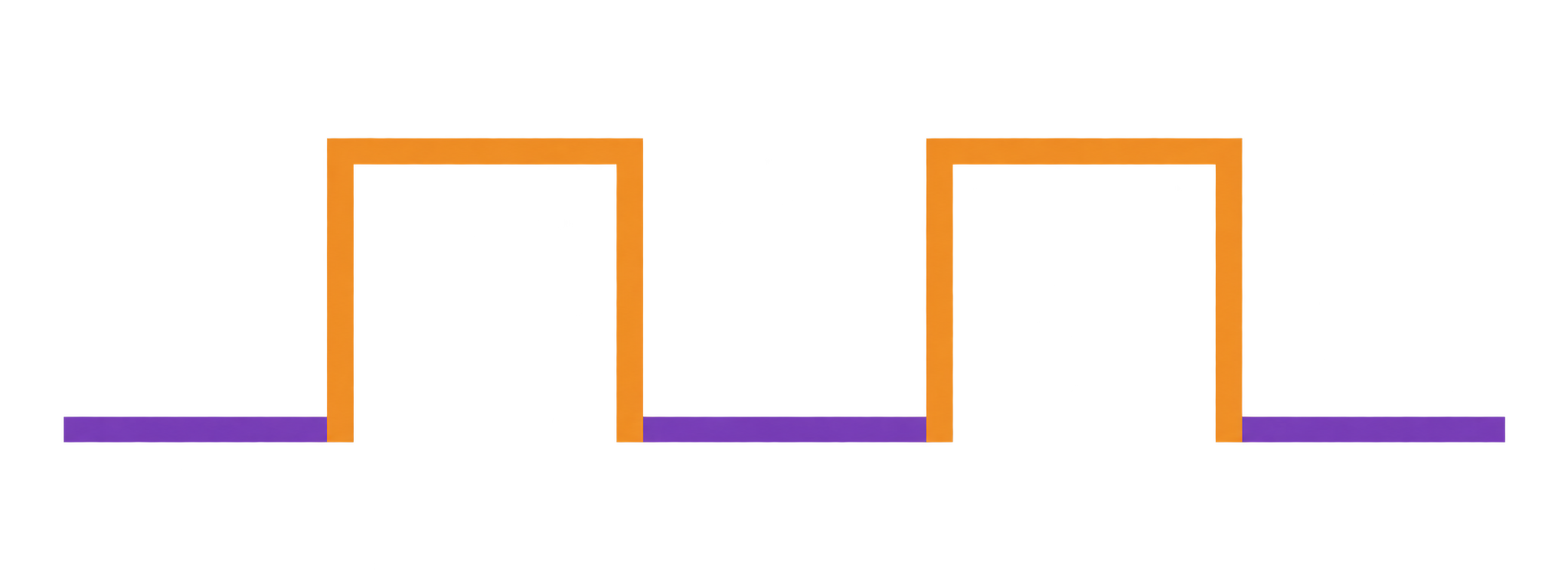}};
\draw [color={rgb, 255:red, 208; green, 2; blue, 27 }  ,draw opacity=1 ]   (432,176) -- (432,212) ;
\draw [color={rgb, 255:red, 208; green, 2; blue, 27 }  ,draw opacity=1 ]   (114,212) -- (432,212) ;
\draw [color={rgb, 255:red, 208; green, 2; blue, 27 }  ,draw opacity=1 ]   (114,179) -- (114,212) ;
\draw [shift={(114,176)}, rotate = 90] [fill={rgb, 255:red, 208; green, 2; blue, 27 }  ,fill opacity=1 ][line width=0.08]  [draw opacity=0] (5.36,-2.57) -- (0,0) -- (5.36,2.57) -- cycle    ;
\draw  [fill={rgb, 255:red, 74; green, 144; blue, 226 }  ,fill opacity=0.5 ] (316,44.38) -- (323.38,37) -- (420,37) -- (420,65.62) -- (412.62,73) -- (316,73) -- cycle ; \draw   (420,37) -- (412.62,44.38) -- (316,44.38) ; \draw   (412.62,44.38) -- (412.62,73) ;
\draw [color={rgb, 255:red, 74; green, 144; blue, 226 }  ,draw opacity=1 ] [dash pattern={on 0.84pt off 2.51pt}]  (300,96) -- (348,96) ;
\draw [color={rgb, 255:red, 74; green, 144; blue, 226 }  ,draw opacity=1 ] [dash pattern={on 0.84pt off 2.51pt}]  (348,76) -- (348,96) ;
\draw [shift={(348,73)}, rotate = 90] [fill={rgb, 255:red, 74; green, 144; blue, 226 }  ,fill opacity=1 ][line width=0.08]  [draw opacity=0] (5.36,-2.57) -- (0,0) -- (5.36,2.57) -- cycle    ;
\draw [color={rgb, 255:red, 74; green, 144; blue, 226 }  ,draw opacity=1 ] [dash pattern={on 0.84pt off 2.51pt}]  (368,76) -- (368,109) ;
\draw [shift={(368,73)}, rotate = 90] [fill={rgb, 255:red, 74; green, 144; blue, 226 }  ,fill opacity=1 ][line width=0.08]  [draw opacity=0] (5.36,-2.57) -- (0,0) -- (5.36,2.57) -- cycle    ;
\draw [color={rgb, 255:red, 74; green, 144; blue, 226 }  ,draw opacity=1 ] [dash pattern={on 0.84pt off 2.51pt}]  (368,109) -- (382,109) ;
\draw [color={rgb, 255:red, 74; green, 144; blue, 226 }  ,draw opacity=1 ] [dash pattern={on 0.84pt off 2.51pt}]  (316,58) -- (114.5,58) ;
\draw [color={rgb, 255:red, 74; green, 144; blue, 226 }  ,draw opacity=1 ] [dash pattern={on 0.84pt off 2.51pt}]  (114.5,58) -- (114.5,84) ;
\draw [shift={(114.5,87)}, rotate = 270] [fill={rgb, 255:red, 74; green, 144; blue, 226 }  ,fill opacity=1 ][line width=0.08]  [draw opacity=0] (5.36,-2.57) -- (0,0) -- (5.36,2.57) -- cycle    ;
\draw (332.25,202) node  {\includegraphics[width=31.88pt,height=16.5pt]{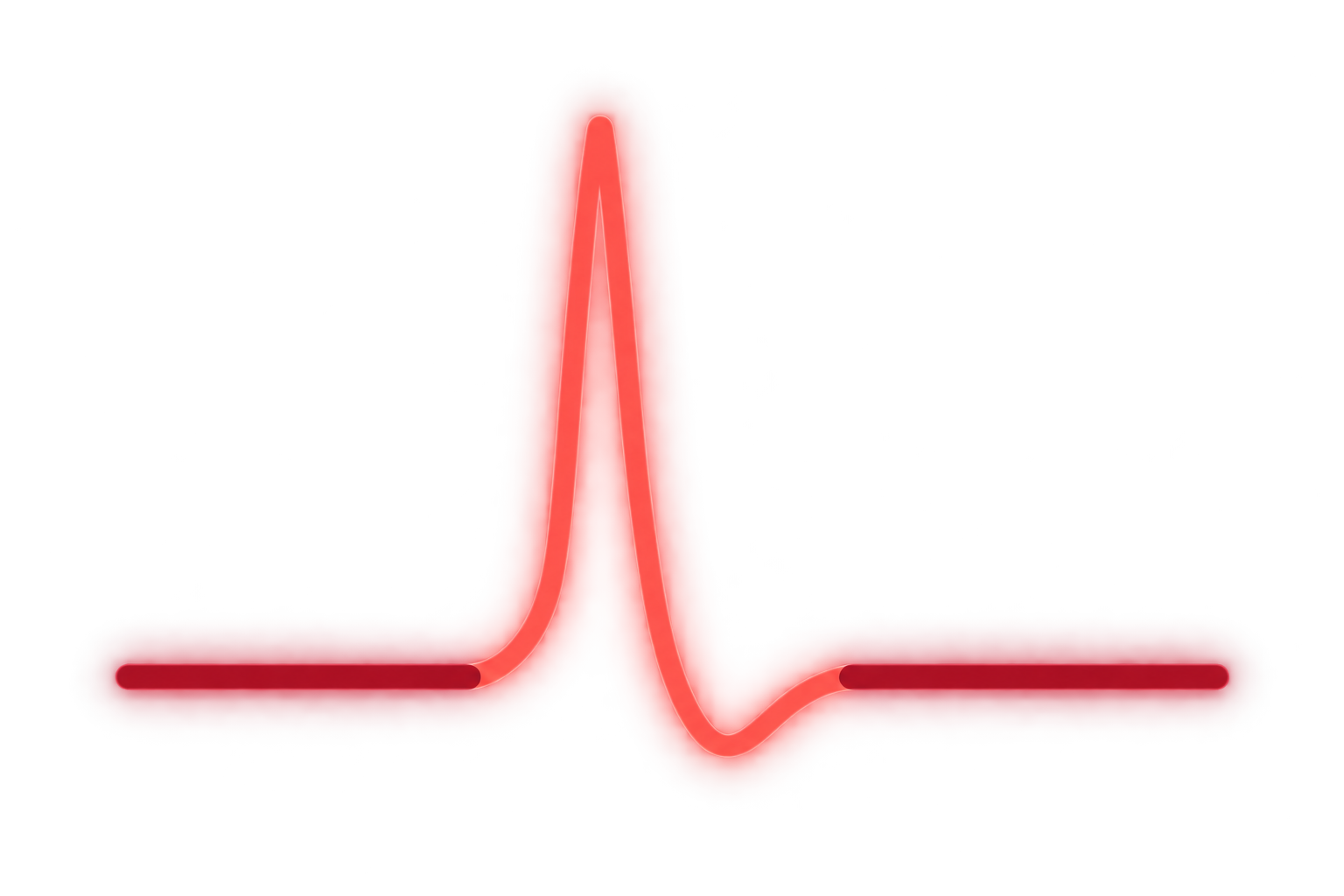}};
\draw (264.75,133) node  {\includegraphics[width=26.63pt,height=14.5pt]{pic_support/spike.png}};
\draw [color={rgb, 255:red, 74; green, 144; blue, 226 }  ,draw opacity=1 ] [dash pattern={on 0.84pt off 2.51pt}]  (301,157) -- (315,157) ;
\draw [color={rgb, 255:red, 74; green, 144; blue, 226 }  ,draw opacity=1 ] [dash pattern={on 0.84pt off 2.51pt}]  (315,105) -- (315,157) ;
\draw [color={rgb, 255:red, 74; green, 144; blue, 226 }  ,draw opacity=1 ] [dash pattern={on 0.84pt off 2.51pt}]  (315,105) -- (359,105) ;
\draw [color={rgb, 255:red, 74; green, 144; blue, 226 }  ,draw opacity=1 ] [dash pattern={on 0.84pt off 2.51pt}]  (358,76) -- (358,105) ;
\draw [shift={(358,73)}, rotate = 90] [fill={rgb, 255:red, 74; green, 144; blue, 226 }  ,fill opacity=1 ][line width=0.08]  [draw opacity=0] (5.36,-2.57) -- (0,0) -- (5.36,2.57) -- cycle    ;

\draw (203,73) node [anchor=north west][inner sep=0.75pt]  [font=\scriptsize] [align=left] {Half-Center Oscillator};
\draw (391,91) node [anchor=north west][inner sep=0.75pt]  [font=\scriptsize] [align=left] {\begin{minipage}[lt]{55.1pt}\setlength\topsep{0pt}
\begin{center}
Scanning Mirror \\System
\end{center}

\end{minipage}};
\draw (215.4,145.29) node [anchor=north west][inner sep=0.75pt]  [font=\scriptsize] [align=left] {Current Source};
\draw (227,125.73) node [anchor=north west][inner sep=0.75pt]  [font=\scriptsize]  {$v_{\mathrm{f}}$};
\draw (74,94) node [anchor=north west][inner sep=0.75pt]  [font=\scriptsize] [align=left] {Predictive Controller};
\draw (176,67.4) node [anchor=north west][inner sep=0.75pt]  [font=\scriptsize]  {$\gsm$};
\draw (173,110.4) node [anchor=north west][inner sep=0.75pt]  [font=\scriptsize]  {$\gus$};
\draw (176,151.4) node [anchor=north west][inner sep=0.75pt]  [font=\scriptsize]  {$I_{0}$};
\draw (344,115.4) node [anchor=north west][inner sep=0.75pt]  [font=\scriptsize]  {$I( t)$};
\draw (53,38) node [anchor=north west][inner sep=0.75pt]  [font=\scriptsize] [align=left] {\begin{minipage}[lt]{113.84pt}\setlength\topsep{0pt}
\begin{center}
Predictive Neuromorphic Controller
\end{center}

\end{minipage}};
\draw (352,197.4) node [anchor=north west][inner sep=0.75pt]  [font=\scriptsize]  {$\El$, $\Er$};
\draw (325,48.5) node [anchor=north west][inner sep=0.75pt]  [font=\scriptsize] [align=left] {\begin{minipage}[lt]{55.9pt}\setlength\topsep{0pt}
\begin{center}
Event-to-Event\\Prediction Model
\end{center}

\end{minipage}};

\end{tikzpicture}

    }
\caption{Predictive Neuromorphic Control Architecture}
\label{fig:architecture}
\end{figure}

Generalized predictive control (GPC) combines model-based prediction with online optimization over a finite horizon~\cite{clarke1987generalized}. This principle also underlies parameterized MPC~\cite{goulart2006optimization}, in which a model predicts the system response under a parameterized policy and optimization selects the policy parameters that best satisfy the control objectives. A neuromorphic controller can similarly be viewed as a dynamic parameterized policy whose internal neuronal states govern actuation. Predicting the coupled evolution of the plant and controller then allows MPC to tune the controller parameters according to their anticipated effects on closed-loop behavior.

Recent developments in parameterized MPC include automatic construction of feedback policies~\cite{jeschke2023grammatical} and learning-based adaptation~\cite{kordabad2023reinforcement}. These formulations rely on predictions of state trajectories, with performance evaluated along the trajectories. They therefore do not directly exploit the information conveyed by sparse sensing events or the temporal organization of neuromorphic actuation. Event-triggered MPC~\cite{brunner2017robust} and the more general asynchronous MPC~\cite{liu2026asynchronous} allow aperiodic feedback updates, but those formulations retain trajectory-based prediction models and control objectives. Closing the control loop at event times therefore does not, by itself, make sensing, prediction, or regulation event-centric.

These observations motivate the development of event-centric MPC (EC-MPC), which places events at the center of sensing, prediction, and control. Events provide feedback information, while an event-to-event (E2E) prediction model predicts their future occurrences. Event timing is then regulated over a horizon defined by a fixed number of future events. This creates a natural synergy between MPC and neuromorphic control, combining model-based prediction and online optimization with sparse actuation and event-based sensing. The contributions of this paper are threefold:

1. We propose the first EC-MPC framework, and we develop a corresponding specified architecture for a scanning mirror system. The architecture uses an E2E prediction model, with events defining the feedback information, prediction horizon, and event-timing objectives.

2. We integrate EC-MPC with a neuromorphic controller to coordinate prescribed zero-crossing times and scanning range while minimizing the energy exchange between the controller and the system.

3. We develop sampling-based algorithms that obtain suboptimal solutions of the EC-MPC optimization problem through implicit evaluation of the E2E prediction model, without requiring an explicit expression for the model.

The rest of the paper is organized as follows. Section~\ref{sec:2-modeling} introduces the scanning mirror dynamics and event sensing. Section~\ref{sec:3-predictive_neuromorphic} presents the neuromorphic control architecture and its parameterization. Section~\ref{sec:4-event_centric_design} develops the EC-MPC architecture, including the E2E prediction model and sampling-based solution algorithms. Section~\ref{sec:5-simulation} presents the numerical evaluation, and Section~\ref{sec:6-conclusion} concludes the paper.

\vspace{0.2cm}
\noindent \textbf{Notation}: The sets of real numbers and nonnegative real numbers are denoted by $\bR$ and $\bR_+$, respectively. The set of natural numbers including $0$ is $\bN$, and $\bI_{a:b} := \bN \cap [a,b]$ for $0 \leq a \leq b$. The matrix or vector p-norm is $\|\cdot\|_p$, where $\|\cdot\|$ defaults to the $2$-norm. The Cartesian product of sets $\cA_1$ and $\cA_2$ is denoted by $\cA_1 \times \cA_2$. For a set $\cA$, its $n$-fold Cartesian product is defined recursively by $\cA^1 := \cA$ and $\cA^n := \cA^{n-1} \times \cA$ for $n \geq 2$. For a column vector $z \in \bR^d$, $z[a:b] := [z_a,\ldots,z_b]^\top$ denotes the subvector from the $a$-th to the $b$-th entry. Given $\bar{x} \in \bR^n$ and a compact set $\cX \subset \bR$, define $\Pi_{\cX}(\bar{x}) := \arg\min_{x\in\cX}\|x - \bar{x}\|$. The $n$-fold vertical stack of $z$ is defined recursively by $\mvec_1\{z\} := z$ and $\mvec_n\{z\} := [(\mvec_{n-1}\{z\})^\top,z^\top]^\top$ for $n \geq 2$. The notation $\mdiag(z)$ denotes the diagonal matrix with diagonal entries $z_1,\ldots,z_d$. For a finite nonempty index set $\mathcal{I} = \{i_1,\ldots,i_m\}$ with a fixed ordering and a family of column vectors $\{z_i\}_{i\in\mathcal{I}}$, we define the vertical stacking operator by $\mvec\{z_i\}_{i\in\mathcal{I}} := [z_{i_1}^\top,\ldots,z_{i_m}^\top]^\top$. For a statement $P$, the indicator $\mathds{1}_{[P]}$ equals $1$ if $P$ is true and $0$ otherwise. The sign function $\operatorname{sgn}:\bR\to\{-1,0,1\}$ is defined by $\operatorname{sgn}(x)=-1$ for $x<0$, $\operatorname{sgn}(x)=0$ for $x=0$, and $\operatorname{sgn}(x)=1$ for $x>0$.
\vspace{-0.1cm}
\section{System Modeling and Event Sensing}
\label{sec:2-modeling}
This section introduces the scanning mirror model and the information available through event sensing. The mechanical dynamics and reluctance actuation mechanism are first described. The photodetector events are then defined.

\subsection{Mechanical Dynamics and Reluctance Actuation}
Consider a single-axis scanning mirror attached to a ferromagnetic mover suspended by a torsional flexure (see Fig.~\ref{fig:mover}). The moving assembly, hereafter referred to as the mover, is modeled as a rigid body symmetric about the $X$, $Y$, and $Z$ axes. The sensing tag is mounted on its longitudinal centerline, which passes through its center of gravity. The mover rotates about a fixed axis (the $X$-axis) that passes through its center of gravity, while the flexure constrains all other degrees of freedom. Within the considered scanning range, the flexure is modeled as a linear torsional spring, and mechanical losses are represented by viscous damping~\cite{pechgraber2023resonant}. The motion is governed by the following Ordinary Differential Equation (ODE):
\begin{equation}
    \label{eq:ode_basic}
    \Mm: \quad
    J\ddot{\alpha} + b\dot{\alpha} + k_\theta\alpha = \tau(\alpha,I),
\end{equation}
where $\alpha \in [-\alpha_{\mathrm{max}},\alpha_{\mathrm{max}}]$ is the angular displacement from the equilibrium position $\alpha =0$, and $\alpha_{\mathrm{max}}>0$ is the maximum allowable deflection angle. The constants $J>0$, $b>0$, and $k_\theta>0$ denote the total moment of inertia of the moving assembly with respect to the $X$-axis, the viscous damping coefficient, and the torsional spring stiffness, respectively. The reluctance torque $\tau(\alpha,I)$ acting on the moving assembly is approximated by
\begin{equation}
    \label{eq:torque_basic}
    \tau(\alpha,I) = -\kappa\sin(\alpha)I^2,
\end{equation}
where $\kappa>0$ is a reluctance torque coefficient. A detailed model of the reluctance force that generates this torque is provided in~\cite{pechgraber2023resonant}. The coil current $I \in [0,I_{\mathrm{max}}]$ serves as the control input, where $I_{\mathrm{max}}>0$ is the maximum allowable current. It is assumed that the current driver is sufficiently fast such that electrical transients can be neglected.
\input{pics/mover_system}

Appropriately timed sparse control inputs can sustain the desired motion while limiting actuator activity and energy consumption. In resonant systems, the natural exchange between kinetic and potential energy supports motion between actuation intervals. Control inputs can therefore be concentrated at suitable phases of the oscillation to compensate for dissipation and regulate its amplitude and timing. To realize such \textit{sparse actuation}, the coil is driven by finite-duration current pulses, with \textit{zero} current between pulses. 
Since the torque is proportional to $I^2$, reversing the current does not reverse the torque. Besides, the reluctance torque always acts toward the equilibrium position ($\alpha = 0$), and its contribution to the mechanical energy depends on the phase of the mirror motion. Specifically, the total mechanical energy
\begin{equation}
    \label{eq:mechanical_energy}
    E = \frac{1}{2}J\dot{\alpha}^2
        + \frac{1}{2}k_\theta\alpha^2
\end{equation}
satisfies
\begin{equation}
    \label{eq:energy_balance}
    \dot{E}
    = -b\dot{\alpha}^2
      -\kappa I^2\sin(\alpha)\dot{\alpha}.
\end{equation}
Therefore, the reluctance torque \textit{supplies} mechanical power when the mirror moves toward the equilibrium position, i.e., $\sin(\alpha)\dot{\alpha}<0$, and \textit{absorbs} mechanical power when the mirror moves away from it, i.e., $\sin(\alpha)\dot{\alpha}>0$. Sustained oscillations require appropriately timed current pulses to compensate for the energy dissipated by damping.

\subsection{Sparse Sensing and Estimation}
The timing of crossings provides information about the motion, allowing feedback control with limited measurement data. Detecting these crossings requires only threshold detection and time stamping, reducing reliance on sensors that continuously resolve position or velocity.

To enable \textit{sparse sensing}, we use two photodetectors. For a fixed incident beam under a planar scanning geometry, the reflected beam is deflected by $2\alpha$ from its nominal direction. The photodetectors, labeled by $\rmL$ and $\rmR$, are positioned at mover directions corresponding to $\alpha=\gamma_{\mathrm{s}}$ and $\alpha=-\gamma_{\mathrm{s}}$, respectively, where $\gamma_{\mathrm{s}}>0$ is chosen close to zero. This sensing configuration supports oscillations at different reference amplitudes $A^\ast \in (\gamma_{\mathrm{s}},\alpha_{\max}]$. Together, the two photodetectors enable estimation of the zero-crossing times and the corresponding angular velocity from the order and timing of their activations. For convenience, we define the photodetector set using their labels as
$\cP=\{\rmL,\rmR\}$.
Each activation of a photodetector $p\in\cP$ is an event denoted by $E_p$. Each event reports the photodetector identity and crossing time, without directly reporting the direction of motion. Let $\Delta_{\mathrm{det}}>0$ denote the maximum admissible time between \textit{paired} photodetector events. A zero-crossing event $\Em$ is inferred when two consecutive events in the combined photodetector event sequence originate from different detectors, either $\El$ followed by $\Er$ or $\Er$ followed by $\El$, and their timestamps $t^-$ and $t^+$ satisfy $0<t^+-t^-\leq\Delta_{\mathrm{det}}$.
\section{Neuromorphic Control Architecture}
\label{sec:3-predictive_neuromorphic}
This section presents the neuromorphic controller and its coupling to the current source. The Half-Center Oscillator (HCO) ~\cite{ribar2021neuromorphic} is first introduced, followed by the mechanism that converts neuronal outputs into current pulses. The tunable controller parameters and the tuning method based on zero-crossing events are then specified.

\subsection{Half-Center Oscillator}
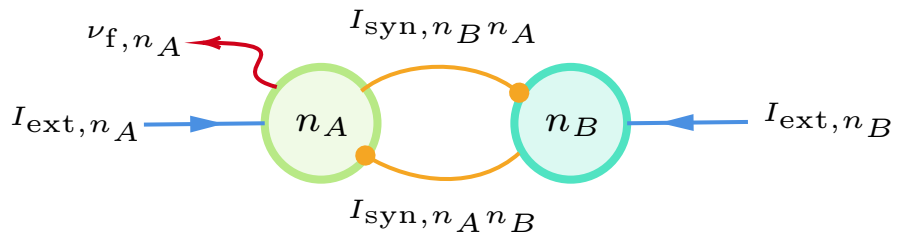
\begin{figure}[h]
    \centering
    \resizebox{0.72\columnwidth}{!}{%

\tikzset{every picture/.style={line width=0.75pt}} 

\begin{tikzpicture}[x=0.75pt,y=0.75pt,yscale=-1,xscale=1]

\draw  [color={rgb, 255:red, 184; green, 233; blue, 134 }  ,draw opacity=1 ][fill={rgb, 255:red, 184; green, 233; blue, 134 }  ,fill opacity=0.2 ][line width=1.5]  (171,95) .. controls (171,87.27) and (177.27,81) .. (185,81) .. controls (192.73,81) and (199,87.27) .. (199,95) .. controls (199,102.73) and (192.73,109) .. (185,109) .. controls (177.27,109) and (171,102.73) .. (171,95) -- cycle ;
\draw  [color={rgb, 255:red, 80; green, 227; blue, 194 }  ,draw opacity=1 ][fill={rgb, 255:red, 80; green, 227; blue, 194 }  ,fill opacity=0.2 ][line width=1.5]  (233,95) .. controls (233,87.27) and (239.27,81) .. (247,81) .. controls (254.73,81) and (261,87.27) .. (261,95) .. controls (261,102.73) and (254.73,109) .. (247,109) .. controls (239.27,109) and (233,102.73) .. (233,95) -- cycle ;
\draw [color={rgb, 255:red, 245; green, 166; blue, 35 }  ,draw opacity=1 ]   (195,87) .. controls (204.2,79.4) and (223.2,78.4) .. (234.2,87.4) ;
\draw [shift={(234.2,87.4)}, rotate = 39.29] [color={rgb, 255:red, 245; green, 166; blue, 35 }  ,draw opacity=1 ][fill={rgb, 255:red, 245; green, 166; blue, 35 }  ,fill opacity=1 ][line width=0.75]      (0, 0) circle [x radius= 2.01, y radius= 2.01]   ;
\draw [color={rgb, 255:red, 245; green, 166; blue, 35 }  ,draw opacity=1 ]   (196,103) .. controls (208.2,110.4) and (224.2,111.4) .. (234.2,102.4) ;
\draw [shift={(196,103)}, rotate = 31.24] [color={rgb, 255:red, 245; green, 166; blue, 35 }  ,draw opacity=1 ][fill={rgb, 255:red, 245; green, 166; blue, 35 }  ,fill opacity=1 ][line width=0.75]      (0, 0) circle [x radius= 2.01, y radius= 2.01]   ;
\draw [color={rgb, 255:red, 74; green, 144; blue, 226 }  ,draw opacity=1 ]   (141,95.33) -- (171,95) ;
\draw [shift={(160.9,95.11)}, rotate = 179.36] [fill={rgb, 255:red, 74; green, 144; blue, 226 }  ,fill opacity=1 ][line width=0.08]  [draw opacity=0] (8.4,-2.1) -- (0,0) -- (8.4,2.1) -- cycle    ;
\draw [color={rgb, 255:red, 74; green, 144; blue, 226 }  ,draw opacity=1 ]   (261,95) -- (291,94.67) ;
\draw [shift={(269.1,94.91)}, rotate = 359.36] [fill={rgb, 255:red, 74; green, 144; blue, 226 }  ,fill opacity=1 ][line width=0.08]  [draw opacity=0] (8.4,-2.1) -- (0,0) -- (8.4,2.1) -- cycle    ;
\draw [color={rgb, 255:red, 208; green, 2; blue, 27 }  ,draw opacity=1 ]   (174,86) .. controls (158.71,82.08) and (183.57,75.28) .. (157.27,75.01) ;
\draw [shift={(155.6,75)}, rotate = 360] [color={rgb, 255:red, 208; green, 2; blue, 27 }  ,draw opacity=1 ][line width=0.75]    (4.37,-1.32) .. controls (2.78,-0.56) and (1.32,-0.12) .. (0,0) .. controls (1.32,0.12) and (2.78,0.56) .. (4.37,1.32)   ;

\draw (177,91.4) node [anchor=north west][inner sep=0.75pt]  [font=\scriptsize]  {$n_{A}$};
\draw (239,91.4) node [anchor=north west][inner sep=0.75pt]  [font=\scriptsize]  {$n_{B}$};
\draw (106,89.4) node [anchor=north west][inner sep=0.75pt]  [font=\tiny]  {$\iext{n_A}$};
\draw (293,88.4) node [anchor=north west][inner sep=0.75pt]  [font=\tiny]  {$\iext{n_B}$};
\draw (190,112.4) node [anchor=north west][inner sep=0.75pt]  [font=\tiny]  {$\isyn{n_A}{n_B}$};
\draw (190,65.4) node [anchor=north west][inner sep=0.75pt]  [font=\tiny]  {$\isyn{n_B}{n_A}$};
\draw (125,69.4) node [anchor=north west][inner sep=0.75pt]  [font=\tiny]  {$\vf{n_A}$};

\end{tikzpicture}

    }
\caption{A half-center oscillator comprising two spiking neurons $n_A$ and $n_B$ interconnected by two synapses (orange curves).}
\label{fig:neuron}
\end{figure}

To generate rhythmic outputs while maintaining a simple controller architecture, we use a neuromorphic controller comprising two spiking neurons that form a half-center oscillator (HCO), as shown in Fig.~\ref{fig:neuron}. The circuit dynamics of each neuron are governed by the following ODEs:
\begin{equation}
\label{eq:ode_hco}
\mathcal{M}_{\mathrm{c}}:\quad
\left\{
\begin{aligned}
\tf\dvf{i} &= -\vf{i} + \gsm\tanh(\vs{i}+0.9) \\
           &\quad - \gus\tanh(\vus{i}+0.9) \\
           &\quad + \isyn{i}{j} + \iext{i},\\
\ts\dvs{i} &= -\vs{i} + \vf{i},\\
\tus\dvus{i} &= -\vus{i} + \vf{i},
\end{aligned}
\right.
\end{equation}
for $i \in \mathcal{N}_{\mathrm{nc}} = \{n_A, n_B\}$, where $j \in \mathcal{N}_{\mathrm{nc}} \setminus \{i\}$ denotes the other neuron. Equations~\eqref{eq:ode_hco} describe the fast, slow, and ultra-slow internal dynamics of each neuron. Here, $\tau_{(\cdot)}$ and $\nu_{(\cdot),i}$ denote the corresponding time constants and voltages, respectively. The term $\isyn{i}{j}$ is the synaptic current from neuron $j$ to neuron $i$, and $\iext{i}$ is the externally applied current. The circuit follows the mixed-feedback model in~\cite{ribar2021neuromorphic} and is configured for rhythmic control using spikes.

The admissible ranges of $\gsm$ and $\gus$ are specified by $[\gsmin,\gsmax]$ and $[\gusmin,\gusmax]$, respectively. These bounds are chosen empirically to ensure that the neurons remain excitable, i.e., capable of generating spikes~\cite{ribar2021neuromorphic}.

\subsection{Spike-Driven Actuation}
The reluctance actuator's current source is directly coupled to either of the two HCO neurons and is active if and only if the neuron's fast voltage reaches or exceeds the threshold $\vth$. Neuron $n_A$ is used by default. The resulting reluctance torque is thus given by
\begin{equation}
    \label{eq:torque_expression_triggering}
    \mathcal{E}_\mathrm{act}:\quad \tau(t) = -\kappa\sin(\alpha(t))[I_0(t)]^2\dsone_{\{\vf{n_A}(t) \geq \vth\}},
\end{equation}
where $I_0$ is the coil current generated by a constant-current source whose output level can be adjusted only intermittently. To accommodate this hardware constraint, the current magnitude is updated only at designated sparse events and held fixed between consecutive events.

\subsection{Optimization-Based Parameter Tuning}
The oscillation frequency and amplitude (i.e., the scanning range of the mirror) can be regulated by tuning the parameters of the excitable neurons~\cite{schmetterling2024neuromorphic} of the HCO and the coil current $I_0$. Specifically, $g_{\mathrm{s}}^-$ and $g_{\mathrm{us}}^+$ influence the spike frequency and spike width, respectively. We collect these tuning parameters into the vector
\begin{equation}
    \theta := [I_0, g_{\mathrm{s}}^-, g_{\mathrm{us}}^+]^\top.
\end{equation}
The admissible parameter set $\Theta \subset \bR^3$ is defined as
\begin{equation}
    \label{eq:parameter_set}
    \Theta = [0, I_{\max}] \times [\gsmin, \gsmax] \times [\gusmin, \gusmax].
\end{equation}
Adaptive tuning of the neuron parameters has been investigated in~\cite{schmetterling2024neuromorphic}. In the current paper, we consider tuning $\theta$ through \textit{event prediction} and \textit{online optimization}, following the working principle of MPC. This approach is closely related to MPC with parameterized control policies, in which the parameters of a \textit{static policy} are optimized online~\cite{goulart2006optimization}. A distinguishing feature of our setting is that the neuromorphic controller has internal states governed by ODEs. Optimizing $\theta$ therefore amounts to tuning a \textit{dynamic controller}. Related approaches include optimization-based tuning of proportional-integral-derivative (PID) controllers~\cite{xu2005auto}, whose integral action introduces an internal state.

We adopt a piecewise-constant parameterization, in which $\theta$ is held constant between consecutive occurrences of $E_{\mathrm{M}}$ and may be updated at each occurrence. In particular, MPC can incorporate predicted changes in references and cost weights alongside predictions of the system behavior. For $n \in \bN$, let $\tm{n}$ denote the time of the $n$-th occurrence of $\Em$ with $\tm{0} = 0$. The policy parameter $\theta$ is held constant over each half-open interval $[\tm{n}, \tm{n+1})$ and is given by
\begin{equation}
\thetae{n} = [\Ie{n}, \gsme{n}, \guse{n}]^\top.
\end{equation}
For each event $\Em$ occurring at $\tm{n}>0$, the two associated consecutive detector events occur in the order $\El$ followed by $\Er$, or $\Er$ followed by $\El$. A schematic overview of the control architecture is given in Fig.~\ref{fig:architecture}. 
\section{Event-Centric Model Predictive Control}
\label{sec:4-event_centric_design}
This section develops the EC-MPC formulation. The E2E prediction model is first constructed, followed by the control objectives and cost function. A finite-horizon optimization problem for EC-MPC is then formulated. Finally, sampling-based algorithms for obtaining suboptimal solutions using the implicit E2E prediction model are presented, together with the online implementation.

\subsection{Event-to-Event Prediction Model}
At time $\tm{n}$, we recursively predict the $j$-step-ahead event times $\htm{j|n}$ for $j \in \bI_{0:N}$, starting from $\htm{0|n}=\tm{n}$. For each $n \in \bI_{0:N-1}$, the following E2E prediction model $\myF: \bR_+\times \Theta \to \bR_+ \times \bR$ determines the next event time $\htm{j+1|n}$ and the swing amplitude $\hbA_{[j,j+1)|n}$ over the corresponding interval $[\htm{j|n},\htm{j+1|n})$:
\begin{equation}
    \label{eq:prediction_model}
    \begin{aligned}
    \begin{bmatrix}
        \htm{j+1|n} \\
        \hbA_{[j,j+1)|n}
    \end{bmatrix}
    &= \myF\bigl(\htm{j|n}, \thetae{j|n}; \\
    &\qquad \dot{\alpha}(\htm{j|n}), \bm{\nu}(\htm{j|n})\bigr),
    \end{aligned}
\end{equation}
where $\thetae{j|n} = [\Ie{j|n}, \gsme{j|n}, \guse{j|n}]^\top$ is the virtual predicted policy parameter and
\begin{equation}
    \label{eq:joint_neuron_state}
    \bm{\nu} := \mvec\bigl\{[\vf{i}, \vs{i}, \vus{i}]^\top\bigr\}_{i \in \mathcal{N}_{\mathbf{nc}}}
\end{equation}
is the state of the neuronal circuit. The E2E prediction model uses the torque expression~\eqref{eq:torque_expression_triggering}, the mechanical dynamics~\eqref{eq:ode_basic}, and the neuromorphic controller dynamics~\eqref{eq:ode_hco}. At each prediction step, it is initialized with $\alpha(\htm{j|n}) = \pm\gamma_{\mathrm{s}}\simeq 0$, the angular velocity $\dot{\alpha}(\htm{j|n})$, and the circuit state at $\htm{j|n}$. Recursively applying the model~\eqref{eq:prediction_model} over $N$ future occurrences of $\Em$ yields the predicted event times $\{\htm{j|n}\}_{j=0}^{N}$.

A closed-form expression for $\myF$ is generally difficult to obtain. Instead, $\htm{j+1|n}$ and $\hbA_{[j,j+1)|n}$ can be evaluated \textit{implicitly} through numerical integration of the coupled mirror-controller dynamics until the next occurrence of $\Em$. The dependence of the predicted event time and scanning range on $\thetae{j|n}$ is therefore implicit.

\begin{remark}
    \label{rmk:interval_varying_parameterization}
    Allowing $\theta$ to vary across intervals provides greater flexibility to accommodate time-varying control objectives and can improve control performance. Keeping $\theta$ constant over the entire prediction horizon, however, reduces the number of decision variables from $3N$ to $3$.
\end{remark}

\subsection{Control Objective \& Cost Function}

The control objective is to regulate the mover such that the event $\Em$ occurs according to a prescribed sequence of desired event times, denoted by $\{\tm{n}^{\ast}\}^{\infty}_{n=0}$ with $\tm{0}^{\ast} = 0$, while the swing amplitude tracks a desired sequence $\{A^\ast_{[n,n+1)}\}^{\infty}_{n=0}$. At the same time, the actuation effort supplied by the motor should be minimized. For convenience, we stack the decision variables as
\begin{equation}
    \label{eq:stacked_theta}
    \boldsymbol\vartheta_{N,j} =[\thetae{0|n}^{\top},\ldots,\thetae{N-1|n}^{\top}]^{\top}.
\end{equation}
Accordingly, at the event instant $\tm{n}$, we consider the following finite-horizon cost function:
\begin{align}
    \label{eq:cost_basic}
        J\left(\boldsymbol\vartheta_N; \dot{\alpha}(\htm{j|n}), \bm{\nu}(\htm{j|n})\right) &\seq
    \sum^{N-1}_{j=0}\gamma^j\Bigg[(\tm{n+j}^\ast - \htm{j|n})^2 \notag \\
    & \hspace{-10ex} + w_{A}(A^\ast_{[n+j,n+j+1)} - \hbA_{[j,j+1)|n})^2 \notag \\
    & \hspace{-10ex} + w_{\tau}\int^{\htm{j+1|n}}_{\htm{j|n}}|\tau(t)\dot{\hA}(t)|\dt\Bigg],
\end{align}
where the discount factor $\gamma \in (0,1]$ assigns progressively less weight to predictions farther into the future, $w_A \in \bR_+$ determines the relative penalty on the amplitude-tracking error, and $w_\tau \in \bR_+$ determines the relative penalty on the mechanical energy transfer between the controller and the mirror system. A special case of interest is periodic oscillation with a constant desired inter-event duration $T$ and a constant desired scanning range $\alpha_0^\ast$, such that
\begin{equation}
\label{eq:periodic_scan_base}
\tm{n}^\ast = nT,
\qquad
|A^\ast_{[n,n+1)}| = A_0^\ast,
\quad \forall n \in \bN.
\end{equation}
Setting $w_A=0$ removes the amplitude-tracking objective, while setting $w_{\tau}=0$ removes the actuator-effort penalty.

\subsection{Finite-Horizon Optimization}
We formulate prediction and receding-horizon optimization in the \textit{event domain}. The event $E_{\rmM}$ triggers a prediction update and a new optimization. The prediction horizon spans the next $N$ occurrences of $E_{\mathrm{M}}$, whose occurrence times are predicted using an E2E prediction model. Thus, the prediction horizon is defined by the number of events rather than a fixed duration. The activation of the photodetectors $\rmL$ and $\rmR$ not only triggers MPC updates, but also provides feedback. In particular, the short interval between consecutive activations of the two photodetectors can be used to infer the zero-crossing angular velocity. This sensing scheme follows the principle of \textit{spike coding}~\cite{ribar2021neuromorphic}, in which information is conveyed through the timing of discrete sensor activations.

The EC-MPC optimization problem for the scanning mirror is then formulated as
\begin{align}
    \label{eq:mpc_optimization}
        \pecmpc:& \min_{\boldsymbol\vartheta_N} J\left(\boldsymbol\vartheta_N; \dot{\alpha}(\htm{0|n}), \bm{\nu}(\htm{0|n})\right) \notag \\
        \mathrm{s.t.}\quad
        & \begin{bmatrix}
            \htm{j+1|n} \\
            \hbA_{[j,j+1)|n}
        \end{bmatrix}
        = \myF\bigl(\htm{j|n}, \thetae{j|n}, \notag \\
        & \hspace{5ex} \dot{\alpha}(\htm{j|n}), \bm{\nu}(\htm{j|n})\bigr),\; \forall n \in \bI_{0:N-1}, \notag \\
        & \boldsymbol\vartheta_N \in \Theta^N, \notag \\
        & \htm{0|n}=\tm{n}, \notag \\
        & \dot{\alpha}(\htm{0|n}) =  \dot{\alpha}^\ast(\htm{0|n}),
\end{align}
where $\dot{\alpha}^\ast(\htm{0|n})$ is the zero-crossing angular velocity measured from the photodetector readings, the cost $J$ is defined in~\eqref{eq:cost_basic}, and $\myF$ is the E2E prediction model in~\eqref{eq:prediction_model}. The optimal value of $\pecmpc$ is denoted by $J^\star(\tm{n})$, and the corresponding set of optimal parameter sequences is denoted by $\Theta^\star_N(\tm{n}) \subset \Theta^N$. 

\subsection{Sampling-Based Solution and Online Implementation}
\label{sec:sampling_solution}
Without a closed-form expression for $\myF$, solving $\pecmpc$ to obtain an analytical characterization of $\Theta^\star_N(\tm{n})$ and a numerical solution of an optimal sequence $\boldsymbol\vartheta_{N,n}^\star \in \Theta^\star_N(\tm{n})$ are generally difficult. We therefore present a tractable rollout-based alternative to obtain a suboptimal parameter sequence $\tilde{\boldsymbol\vartheta}^\star_{N,n}$.

\subsubsection{Joint-dynamics rollouts}
For each candidate parameter sequence, we integrate the mirror and the neuronal circuit together. Define the state of the joint system by $x=[\alpha,\omega,\bm{\nu}^{\top}]^{\top}\in\bR^8$, where $\omega=\dot\alpha$ and $\bm\nu$ is given in~\eqref{eq:joint_neuron_state}. On the $n$-th predicted inter-event interval, the joint dynamics are
\begin{equation}
\label{eq:joint_rollout}
\begin{aligned}
 \dot{\hat\alpha}&=\hat\omega,\\
 J\dot{\hat\omega}&=-b\hat\omega-k_\theta\hat\alpha
 -\kappa\sin(\hat\alpha)\Ie{j|n}^{2}\hat\chi,\\
 \dot{\hat{\bm\nu}}&=f_{\mathrm c}(\hat{\bm\nu};\gsme{j|n},\guse{j|n}),\\
 \hat\chi&=\dsone_{\{\hat\nu_{\mathrm f,n_A}\geq\vth\}},
\end{aligned}
\end{equation}
where $f_{\mathrm c}$ is the vector field in~\eqref{eq:ode_hco}, including the synaptic currents. Here, $\Ie{j|n}$ is the commanded current amplitude; the actual coil current is $\Ie{j|n}\hat\chi$. Every candidate starts from the same state estimate
\begin{equation}
    \label{eq:roll_out_initial_state}
    \hat x(\tm{n})=[0,\dot\alpha^\ast(\tm{n}),\bm\nu(\tm{n})^{\top}]^{\top}.
\end{equation}
The numerical solver locates the times at which the condition $\hat\nu_{\mathrm f,n_A}-\vth$ holds, thereby determining the onset and termination of each current pulse. The integrator locates the next nontrivial zero of $\hat\alpha$, after the mirror has left its current crossing, and records it as $\htm{j+1|n}$. For a half-swing with one turning point, define the signed extremum by
\begin{equation}
\label{eq:rollout_amplitude}
 \hbA_{[j,j+1)|n}
 =\sigma_{j|n}\max_{t\in[\htm{j|n},\htm{j+1|n}]}|\hat\alpha(t)|,
\end{equation}
where $\sigma_{j|n}=\operatorname{sgn}(\hat\omega(\htm{j|n}))$; the desired extrema use the same sign convention. Turning points are located from $\hat\omega=0$. The energy-exchange term is evaluated by adding the auxiliary state $q$ whose dynamics is given by
\begin{equation}
 \dot q=|\hat\tau\hat\omega|,\qquad
 q(\htm{j|n})=0.
\end{equation}
The terminal value of $q$ gives the integral in~\eqref{eq:cost_basic}. At each predicted zero-crossing, only $\theta$ and $p$ change; the mechanical velocity and all circuit states are carried forward continuously. The rollout ends after $N$ future crossings; thus, the time horizon depends on the candidate sequence.

\begin{remark}
    Circuit states are retained between MPC updates and are read from the numerical controller or propagated from a known initialization.
\end{remark}

\begin{remark}
    A candidate is rejected if (i) the angle exceeds a predefined maximum range (i.e., $|\hat\alpha|>\alpha_{\max}$), (ii) a half-swing fails to reach the photodetectors, or (iii) no subsequent transverse zero crossing occurs within the prescribed timeout $H_{\max}$. These checks enforce the operating regime of the E2E model and limit the simulated duration of each rollout.
\end{remark}

\subsubsection{Sampling method}
We use a parameter-space search inspired by model predictive path integral (MPPI) control~\cite{williams2017model} to sample rollouts. For event index $n$, at iteration $r$, sample perturbations around a nominal sequence $\bar{\boldsymbol\vartheta}^{(r)}_n$ and project onto the admissible parameter set:
\begin{equation}
\label{eq:parameter_sampling}
\begin{aligned}
 \boldsymbol\vartheta^{(m,r)}_n
 &=\Pi_{\Theta^N}\bigl(\bar{\boldsymbol\vartheta}^{(r)}_n
                 +\varepsilon^{(m,r)}\bigr),\\
 \varepsilon^{(m,r)}&\sim\mathcal N(0,\Sigma),
 \qquad m=1,\ldots,M-1.
\end{aligned}
\end{equation}
For $m=0$, we set $\boldsymbol\vartheta^{(0,r)}_n = \bar{\boldsymbol\vartheta}^{(r)}_n$. Each candidate is evaluated by a joint rollout using the cost function~\eqref{eq:cost_basic}. Denote its cost as $J^{(m,r)}_n$, and let $\mathcal{V}^{(r)}_n$ be the set that contains the indices of valid rollouts. For $\mathcal{V}^{(r)}_n\neq\emptyset$, define
\begin{subequations}
\label{eq:mppi_weights}
\begin{align}
 \rho^{(r)}_n&:=\min_{m\in\mathcal{V}^{(r)}_n}J^{(m,r)}_n,\\
 w^{(m,r)}_n&:=\frac{\exp[-(J^{(m,r)}_n-\rho^{(r)}_n)/\lambda]}
 {\sum_{\ell\in\mathcal{V}^{(r)}_n}\exp[-(J^{(\ell,r)}_n-\rho^{(r)}_n)/\lambda]},
 \quad m\in\mathcal \mathcal{V}^{(r)}_n,
\end{align}
\end{subequations}
where $\lambda>0$ is the scaling parameter\footnote{The parameter $\lambda$ is commonly referred to as the temperature parameter.} that controls the sensitivity of the weights to cost variation. Smaller $\lambda$ concentrates the weights on candidates with lower costs, whereas larger values distribute the weights more uniformly across valid candidates. The next nominal sequence is given by
\begin{equation}
\label{eq:mppi_update}
 \bar{\boldsymbol\vartheta}^{(r+1)}_n
 =\sum_{m\in\mathcal{V}^{(r)}_n}w^{(m,r)}_n\boldsymbol\vartheta^{(m,r)}_n.
\end{equation}
By convexity of $\Theta$, $\bar{\boldsymbol\vartheta}^{(r+1)}_n \in \Theta^N$. After $R$ rounds of iterations ($r = 1,2,\dots, R$), the valid parameter sequence with the lowest cost among all evaluated candidates, including the evaluated mean sequences, is returned as $\tilde{\boldsymbol\vartheta}^\star_{N,n}$.
\begin{remark}
    The sampling strategy in~\eqref{eq:parameter_sampling} comprises $R$ sequential rounds, with the best candidate from each round guiding sampling in the next. Although these rounds cannot be parallelized, this progressive refinement can reduce the total number of samples needed to obtain a satisfactory solution. Alternatively, setting $R=1$ yields a random-shooting method that draws a single batch uniformly from $\Theta^N$ and returns the best candidate. This approach is fully parallelizable but may require substantially more samples to achieve comparable performance. Imposing $\thetae{0\vert{}n} = \cdots = \thetae{N-1\vert{}n}$ reduces the search dimension and can further reduce the required sample budget, recovering the constant-parameter option discussed in Remark~\ref{rmk:interval_varying_parameterization}. The present paper does not establish a certified bound on the optimality gap of the proposed MPPI-inspired solver for $\pecmpc$ in~\eqref{eq:mpc_optimization}.
\end{remark}

\subsubsection{Receding-horizon implementation}
At the next event, the current solution is shifted by one event interval with its last entry repeated to initialize the sampling, i.e.,
\begin{equation}
    \label{eq:sample_assign}
    \hspace{-1ex}\bar{\boldsymbol\vartheta}^{(0)}_{n+1} \seq \left[\bigl(\tilde{\boldsymbol\vartheta}^\star_{N,n}[4:3N]\bigr)^\top, \bigl(\tilde{\boldsymbol\vartheta}^\star_{N,n}[3N\sminus2:3N]\bigr)^\top\right]^\top.
\end{equation}
After obtaining $\tilde{\boldsymbol\vartheta}^\star_{N,n}$, only $\tthetae{0|n}$ is applied; the HCO then generates the pulses until the next event update. Candidate rollouts are independent and can be evaluated in parallel. If no valid sequence is found, a defined failure action is to set the pulse amplitude to zero and retain the circuit gains; this action does not guarantee sustained oscillation.

For two detection times $t_{n}^-<t_{n}^+$ associated with the same passage, the feedback estimates are
\begin{subequations}
\label{eq:event_feedback_estimates}
\begin{align}
&\tm{n}\simeq\frac{t_{n}^-+t_{n}^+}{2}\simeq t_{n}^-\simeq t_{n}^+, \\
 &\dot\alpha^\ast(\tm{n})\simeq
 \frac{2\sigma_n\gamma_{\mathrm s}}{t_{n}^+-t_{n}^-},
\end{align}
\end{subequations}
where $\sigma_n=+1$ for $\Er$ followed by $\El$, and $\sigma_n=-1$ for the reverse order. These estimates become available only at $t_n^+$. The small value of $\gamma_{\mathrm{s}}$ justifies the estimates in~\eqref{eq:event_feedback_estimates}. A causal implementation must retain the previous parameter values until the new solution becomes available and propagate both the mechanical and circuit states to the corresponding update time.
\section{Numerical Simulation}
\label{sec:5-simulation}
This section evaluates the proposed EC-MPC framework through numerical simulations. First, the simulation setup is described, then the control performance is presented. Finally, the issues of computational cost and model mismatch are discussed. The simulations\footnote{The full codebase and specific implementations are available at: \url{https://github.com/lcrekko/event_centric_mpc}.} are implemented in Python 3.12.

\subsection{Simulation Setup}
\label{subsec:5-1-simulation_setup}
We consider a numerical example combining the scanning mirror of Section~\ref{sec:2-modeling} with the HCO of Section~\ref{sec:3-predictive_neuromorphic}. The reference oscillation frequency is $f_{\mathrm{ref}}=10\,\mathrm{Hz}$, corresponding to a desired inter-event duration $T=1/(2f_{\mathrm{ref}})=50\,\mathrm{ms}$.

\subsubsection{Mechanical system}
Table~\ref{tab:physical_parameters} summarizes the mechanical parameters and operating limits. The mover is characterized by $J$, $k_\theta$,
and $b$. These parameters yield
a zero-current natural frequency and damping ratio of
\begin{equation}
 f_0=\frac{1}{2\pi}\sqrt{\frac{k_\theta}{J}}
 =9.9\,\mathrm{Hz},\qquad
 \zeta=\frac{b}{2\sqrt{Jk_\theta}}=0.01.
\end{equation}
During current pulses, the electromagnetic torque provides
an additional restoring contribution, increasing the effective
stiffness near equilibrium. The torque coefficient $\kappa=0.03\,\mathrm{N\,m/A^2}$ is obtained based on the small-angle coefficient $0.54\,\mathrm{mN\,m/(deg\,A^2)}$ of~\cite{pechgraber2023resonant}. The simulation retains the $\sin\alpha$ dependence in~\eqref{eq:torque_basic}. The current limit is $I_{\max}=0.12\,\mathrm A$ to match the mechanical stiffness.
\begin{table}[h]
\centering
\caption{Mechanical parameters and operating limits.}
\label{tab:physical_parameters}
\small
\setlength{\tabcolsep}{3pt}
\renewcommand{\arraystretch}{1.12}
\rowcolors{2}{simTableStripe}{white}
\begin{tabularx}{\columnwidth}{@{}lXlX@{}}
\toprule
\rowcolor{simTableHead}
Parameter & Value & Parameter & Value\\
\midrule
$J$ & $10^{-7}\,\mathrm{kg\,m^2}$ & $\zeta$ & $0.01$\\
$f_0$ & $9.9\,\mathrm{Hz}$ & $\kappa$ & $30.85\,\mathrm{mN\,m/A^2}$\\
$k_\theta$ & $0.39\,\mathrm{mN\,m/rad}$ &
$b$ & $0.12\,\mu\mathrm{N\,m\,s/rad}$\\
$I_{\max}$ & $0.12\,\mathrm A$ & $\alpha_{\max}$ & $\pi/18$\\
$\gamma_{\mathrm s}$ & $\pi/3600$ & $A_0^\ast$ & $\pi/36$\\
\bottomrule
\end{tabularx}
\end{table}

\subsubsection{Neuronal circuit}
The two neurons are identical and connected by reciprocal \textit{inhibitory} synapses. The synaptic current flowing into neuron $i \in \mathcal{N}_{\mathrm{nc}}$ is given by~\cite{schmetterling2024neuromorphic}
\begin{equation}
\label{eq:simulation_synapse}
 \isyn{i}{j}=
 \frac{g_{\mathrm{syn}}}{1+\exp[-2(\vs{j}+1)]},
\end{equation}
where $j \in \mathcal{N}_{\mathrm{nc}}\setminus\{i\}$, and $g_{\mathrm{syn}} < 0$ is the synaptic gain. In the simulation comparison, we start from the default neuronal gains $(\gsm,\gus)=(1.5,2)$ to show that EC-MPC can meet the control objectives without offline tuning based on human expertise or extensive trial and error. The simulated HCO has an estimated dimensionless recurrence period $P_{\mathrm H}\simeq286.32$ between successive upward threshold crossings corresponding to neuron $n_A$, and the time constants are then given by
\begin{equation}
\label{eq:neuron_time_scaling}
 c_{\mathrm t}=\frac{T}{P_{\mathrm H}}
 \simeq174.63\,\mu\mathrm s,\;\;
 (\tf,\ts,\tus)=(1,10,100)c_{\mathrm t}.
\end{equation}
The nominal current pulse has a period of $T$, generating two pulses per mechanical period. The duration of each current pulse is approximately $13.58\,\mathrm{ms}$. The time constants remain fixed when the gains are optimized online. 
\begin{table}[h]
\centering
\caption{Nominal neuronal parameters.}
\label{tab:neuronal_parameters}
\small
\setlength{\tabcolsep}{3pt}
\renewcommand{\arraystretch}{1.12}
\rowcolors{2}{simTableStripe}{white}
\begin{tabularx}{\columnwidth}{@{}lXlX@{}}
\toprule
\rowcolor{simTableHead}
Parameter & Value & Parameter & Value\\
\midrule
$g_{\mathrm{syn}}$ & $-1$ & $\iext{i}$ & $-1$\\
$\gsm$ & $1.5$ & $\gus$ & $2$\\
$\vth$ & $0$ & $\tf$ & $0.17\,\mathrm{ms}$\\
$\ts$ & $1.75\,\mathrm{ms}$ & $\tus$ & $17.46\,\mathrm{ms}$\\
\bottomrule
\end{tabularx}
\end{table}

\vspace{-0.3cm}
\subsubsection{Initialization and event feedback}
The reference event times and signed half-swing extrema are chosen to support periodic scanning (cf.~\eqref{eq:periodic_scan_base}):
\begin{equation}
\label{eq:periodic_scan_simulation}
 \tm{j}^{\ast}=jT,\quad
 A^{\ast}_{[j,j+1)}=(-1)^jA_0^\ast,\quad
 A_0^\ast=\pi/36.
\end{equation}
The system starts with an established oscillation with $\alpha(0)=0$, $\dot\alpha(0)=5.51\,\mathrm{rad/s}$, and $I_0\simeq50\,\mathrm{mA}$. The initial circuit state is given by $\bm\nu_{n_A}(0)\simeq[-0.77,-0.44,-0.45]^\top$ and $\bm\nu_{n_B}(0)\simeq[-2.32,-2.43,-2.15]^\top$. The current pulses are expected to act during the return toward the equilibrium ($\alpha = 0$), supplying energy (cf.~\eqref{eq:energy_balance}) to sustain the oscillation in the presence of damping.

The photodetectors are placed at $\alpha=\pm\gamma_{\mathrm s}$. Opposite-detector activations are paired when their separation is below $\Delta_{\mathrm{pair}}=0.05T=2.5\,\mathrm{ms}$, and the feedback estimates are computed using~\eqref{eq:event_feedback_estimates}. We assume noiseless timestamps and negligible sensing and computation delays, as discussed in Section~\ref{sec:sampling_solution}. The controller uses only the event-derived mechanical estimate and its retained circuit state.

\subsubsection{MPC settings}
The standard deviations for sampling are $(\sigma_I,\sigma_{\mathrm s},\sigma_{\mathrm{us}})$, which form the covariance in~\eqref{eq:parameter_sampling} as $\Sigma = \mdiag(\mvec_N([\sigma_I^2,\sigma_{\mathrm s}^2,\sigma_{\mathrm{us}}^2]^\top))$. The horizon for MPC spans $N=4$ half-swings, with a time horizon of $NT=0.2\,\mathrm s$.
\begin{table}[h]
\centering
\caption{model predictive control parameters.}
\label{tab:solver_parameters}
\small
\setlength{\tabcolsep}{3pt}
\renewcommand{\arraystretch}{1.12}
\rowcolors{2}{simTableStripe}{white}
\begin{tabularx}{\columnwidth}{@{}lXlX@{}}
\toprule
\rowcolor{simTableHead}
Parameter & Value & Parameter & Value\\
\midrule
$N$ & $4$ & $M$ & $256$\\
$R$ & $2$ & $\gamma$ & $1$\\
$[\gsmin,\gsmax]$ & $[1.5,2.5]$ &
$[\gusmin,\gusmax]$ & $[1,2]$\\
$\sigma_I$ & $12\,\mathrm{mA}$ &
$(\sigma_{\mathrm s},\sigma_{\mathrm{us}})$ & $(0.04,0.04)$\\
$\lambda$ & $250\,\mathrm{ms^2}$ &
$H_{\max}$ & $100\,\mathrm{ms}$\\
$\epsilon_{\mathrm{rel}}$ & $10^{-7}$ &
$\epsilon_{\mathrm{abs}}$ & $10^{-9}$\\
$h_{\max}$ & $87.32\,\mu\mathrm s$ &
$K$ & $200$\\
$s_{\mathrm{rng}}$ & $1$ & $\Delta_{\mathrm{pair}}$ & $2.5\,\mathrm{ms}$\\
\bottomrule
\end{tabularx}
\end{table}

The nominal kinetic energy of the mover system is given by $E_0=\tfrac12 k_\theta(\alpha_0^\ast)^2\simeq1.47\,\mu\mathrm J$, and the cost weights are then chosen as
\begin{equation}
\label{eq:simulation_weights}
\begin{aligned}
 w_A&=\frac{T^2}{(\alpha_0^\ast)^2}
 \simeq0.33\,\mathrm{s^2/rad^2},\\
 w_\tau&=0.05\frac{T^2}{E_0}
 \simeq84.84\,\mathrm{s^2/J}.
\end{aligned}
\end{equation}

To resolve rapid neuronal transients while allowing larger steps when the solution varies slowly, we perform numerical rollouts using an adaptive fourth-order Runge--Kutta method~\cite{hairer1993solving} with relative tolerance $\epsilon_{\mathrm{rel}}$, scaled absolute tolerance $\epsilon_{\mathrm{abs}}$, and maximum step size $h_{\max}=\tf/2$. 
The plant and prediction model use the same nominal parameters. The number of simulated inter-event intervals is $K=50$, corresponding to approximately $2.5\,\mathrm s$.
\subsection{Results and Performance Evaluation}
\begin{figure}[htbp]
    \centering
    \includegraphics[width=\columnwidth]{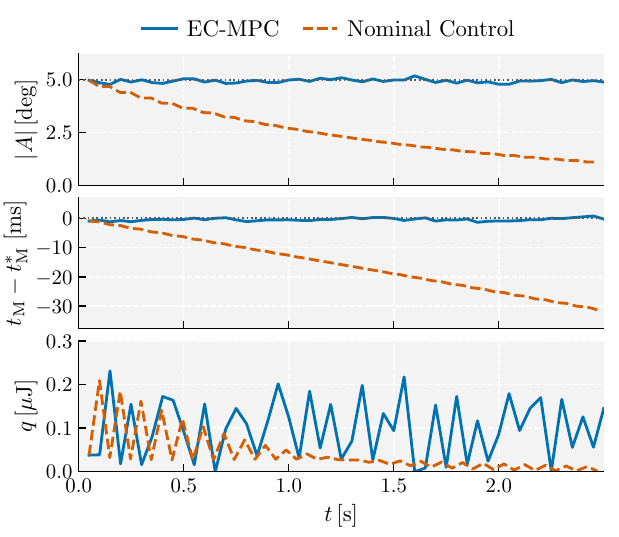}
    \caption{Scanning range tracking, zero-crossing timing, and mechanical energy exchange under EC-MPC and the nominal controller with fixed default parameters.}
    \label{fig:performance}
\end{figure}

We compare the EC-MPC controller with a nominal neuromorphic controller using the fixed default parameter $\theta_{\mathrm{df}} = [50\mathrm{mA}, 1.5, 2]^\top$. Initialized with the same parameter $\theta_{\mathrm{df}}$, the EC-MPC controller subsequently updates its parameters through online predictive optimization. This comparison highlights the benefits of predictive parameter tuning, which reduces reliance on offline tuning. Figure~\ref{fig:performance} compares their performance in terms of tracking scanning range, tracking zero-crossing times, and mechanical energy transfer. The results show that EC-MPC sustains oscillations at the desired amplitude while tracking the prescribed zero-crossing times. The energy transfer compensates for the energy dissipated by damping. Without feedback-based parameter tuning, the nominal controller exhibits worse tracking performance and eventually fails to supply energy to sustain the oscillations.

\begin{figure}[ht]
    \centering
    \includegraphics[width=\columnwidth]{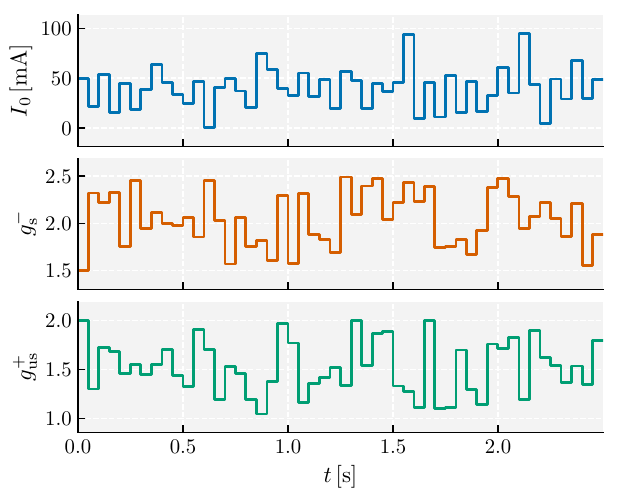}
    \caption{Evolution of the controller parameters updated by EC-MPC at zero-crossing events.}
    \label{fig:parameter}
\end{figure}

Figure~\ref{fig:parameter} shows the controller parameters updated by EC-MPC at each zero-crossing event. Although satisfactory tracking is achieved, the parameters do not exhibit clear convergence to fixed values. These variations may reflect the stochastic nature of the sampling-based solver, which uses a moderate sample size and a small number of refinement rounds. They may also arise because different parameter combinations can yield comparable control performance, with the solver selecting a single candidate at each update.

\begin{figure}[htbp]
    \centering
    \includegraphics[width=\columnwidth]{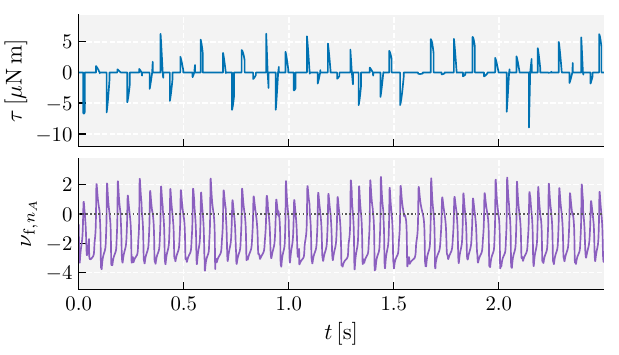}
    \caption{Applied torque and fast voltage of neuron $n_A$, illustrating actuator activation by neuronal spikes.}
    \label{fig:actuation}
\end{figure}

\vspace{-0.13cm}
Figure~\ref{fig:actuation} shows the torque applied to the mover and the fast voltage of neuron $n_A$, which provides the actuator activation signal. This correspondence illustrates how the neuromorphic controller implements sparse actuation through brief mechanical interventions.

\subsection{Discussion}
The computational cost of rollouts limits the oscillation frequencies at which EC-MPC can operate in real time, as the optimization must finish before the next event. Parallel sample evaluation on multicore processors or graphical processing units (GPUs) can reduce computation time and permit larger sample budgets, potentially improving solution quality. However, the number of sequential refinement rounds within each MPC optimization should remain limited.

Although trajectory rollouts avoid the need for a closed-form E2E prediction model, they still require ODE-based models of both the plant and the neuronal circuits. EC-MPC thus remains sensitive to model mismatch, whose effects may accumulate over the prediction window. Shorter event horizons may be preferable under substantial mismatch. Both the impact of model mismatch and the joint influence of model mismatch and the prediction horizon on control performance warrant further investigation in EC-MPC, building on existing studies of conventional MPC~\cite{liu2024stability, liu2025regret, liu2026certainty,liu2026asynchronous}.

Finally, EC-MPC only supports enforcing state constraints (i.e., $|\alpha| \sleq \alpha_{\max}$) at event times, without guarantees on constraint satisfaction between events. Discrete-time control barrier functions~\cite{liu2027robust} can help address this issue, following a mode-based approach where the inter-event dynamics and potential modeling errors are accounted for.
\vspace{-0.05cm}
\section{Conclusions and Future Work}
\label{sec:6-conclusion}
This work has developed an event-centric model predictive control (EC-MPC) framework and a corresponding control architecture for a scanning mirror system. By defining the prediction horizon through future events and optimizing neuromorphic controller parameters at event instants, EC-MPC integrates prediction and online optimization with sparse sensing and actuation. Numerical simulations demonstrate that the proposed architecture sustains oscillations at the desired amplitude while tracking prescribed zero-crossing times. The improved performance over a nominal controller with fixed parameters highlights the value of feedback-based parameter tuning and demonstrates the potential of predictive optimization for neuromorphic control.

Future work will investigate constructing event-to-event prediction models from offline rollouts to reduce online computational cost while retaining online optimization. Further directions include quantifying the effects of plant-model mismatch, as well as incorporating control barrier functions to provide safety guarantees between consecutive events.


\vspace{-0.05cm}
\bibliographystyle{IEEEtran}
\bibliography{references/cbf,references/mpc,references/spiky}

@article{liu2027robust,
  title={Robust adaptive discrete-time control barrier certificate},
  author={C. Liu and A. Alan and S. Shi and B. {D}e Schutter},
  journal={Automatica},
  volume={195},
  pages={113310},
  year={2027},
  publisher={Elsevier}
}

@article{liu2025regret,
	title={On the Regret of Model Predictive Control With Imperfect Inputs},
	author={C. Liu and S. Shi and B. {D}e Schutter},
	journal={IEEE Control Systems Letters},
	pages={601--606},
	volume={9},
	year={2025},
	publisher={IEEE}
}

@article{liu2026certainty,
	title={Certainty-equivalence model predictive control: {S}tability, performance, and beyond},
	author={C. Liu and S. Shi and B. {D}e Schutter},
	journal={IEEE Transactions on Automatic Control},
    volume={71},
    number={7},
    pages={4649--4664},
	year={2026},
	publisher={IEEE}
}

@article{goulart2006optimization,
  title={Optimization over state feedback policies for robust control with constraints},
  author={P. J. Goulart and E. C. Kerrigan and J. M. Maciejowski},
  journal={Automatica},
  volume={42},
  number={4},
  pages={523--533},
  year={2006},
  publisher={Elsevier}
}

@article{xu2005auto,
  title={Auto-tuning of {PID} controller parameters with supervised receding horizon optimization},
  author={M. Xu and S. Li and C. Qi and W. Cai},
  journal={ISA transactions},
  volume={44},
  number={4},
  pages={491--500},
  year={2005},
  publisher={Elsevier}
}

@article{clarke1987generalized,
  title={Generalized predictive control—{P}art {I}. {T}he basic algorithm},
  author={D. W. Clarke and C. Mohtadi and P. S. Tuffs},
  journal={Automatica},
  volume={23},
  number={2},
  pages={137--148},
  year={1987},
  publisher={Elsevier}
}

@article{kordabad2023reinforcement,
  title={Reinforcement learning for {MPC}: {F}undamentals and current challenges},
  author={A. B. Kordabad and D. Reinhardt and A. S. Anand and S. Gros},
  journal={IFAC-PapersOnLine},
  volume={56},
  number={2},
  pages={5773--5780},
  year={2023},
  publisher={Elsevier}
}

@article{brunner2017robust,
  title={Robust event-triggered {MPC} with guaranteed asymptotic bound and average sampling rate},
  author={F. D. Brunner and W. P. M. H. Heemels and F. Allg{\"o}wer},
  journal={IEEE Transactions on Automatic Control},
  volume={62},
  number={11},
  pages={5694--5709},
  year={2017},
  publisher={IEEE}
}

@article{liu2026asynchronous,
  title={Asynchronous model predictive control under model mismatch: {S}tability and performance guarantees},
  author={C. Liu and A. Alan and S. Shi and B. {D}e Schutter},
  journal={arXiv preprint arXiv:2609.08836},
  year={2026}
}

@article{jeschke2023grammatical,
  title={Grammatical-evolution-based parameterized model predictive control for urban traffic networks},
  author={J. Jeschke and D. Sun and A. Jamshidnejad and B. {D}e Schutter},
  journal={Control Engineering Practice},
  volume={132},
  pages={105431},
  year={2023},
  publisher={Elsevier}
}

@article{williams2017model,
  title={Model predictive path integral control: From theory to parallel computation},
  author={G. Williams and A. Aldrich and E. A. Theodorou},
  journal={Journal of Guidance, Control, and Dynamics},
  volume={40},
  number={2},
  pages={344--357},
  year={2017},
  publisher={American Institute of Aeronautics and Astronautics}
}

@inproceedings{liu2024stability,
  title={Stability and performance analysis of model predictive control of uncertain linear systems},
  author={C. Liu and S. Shi and B. {D}e Schutter},
  booktitle={2024 IEEE 63rd Conference on Decision and Control (CDC)},
  pages={7356--7362},
  year={2024},
}

@IEEEtranBSTCTL{BSTcontrol,
  CTLdash_repeated_names = "no"
}

@article{arbelaiz2025spiking,
  title={Spiking control systems for soft robotics: a rhythmic case study in a soft robotic crawler},
  author={J. Arbelaiz and A. Franci and N. E. Leonard and R. Sepulchre and B. Bamieh},
  journal={arXiv preprint arXiv:2509.02968},
  year={2025}
}

@article{schmetterling2024neuromorphic,
  title={Neuromorphic control of a pendulum},
  author={R. Schmetterling and F. Forni and A. Franci and R. Sepulchre},
  journal={IEEE Control Systems Letters},
  volume={8},
  pages={1235--1240},
  year={2024},
  publisher={IEEE}
}

@article{pechgraber2023resonant,
  title={Resonant rotational reluctance actuator for large range scanning mirrors},
  author={D. Pechgraber and E. Csencsics and G. Schitter},
  journal={IEEE/ASME Transactions on Mechatronics},
  volume={28},
  number={6},
  pages={3573--3582},
  year={2023},
  publisher={IEEE}
}

@article{ribar2021neuromorphic,
  title={Neuromorphic control: {D}esigning multiscale mixed-feedback systems},
  author={L. Ribar and R. Sepulchre},
  journal={IEEE Control Systems Magazine},
  volume={41},
  number={6},
  pages={34--63},
  year={2021},
  publisher={IEEE}
}

@article{ji2008advances,
  title={Advances in the speed and resolution of light microscopy},
  author={N. Ji and H. Shroff and H. Zhong and E. Betzig},
  journal={Current Opinion in Neurobiology},
  volume={18},
  number={6},
  pages={605--616},
  year={2008},
  publisher={Elsevier}
}

@article{de2013polygon,
  title={Polygon scanner system for ultra-short pulsed laser micro-machining applications},
  author={R. {D}e Loor},
  journal={Physics Procedia},
  volume={41},
  pages={544--551},
  year={2013},
  publisher={Elsevier}
}

@inproceedings{kimoto2014development,
  title={Development of small size 3{D} {L}i{DAR}},
  author={K. Kimoto and N. Asada and T. Mori and Y. Hara and A. Ohya and S. Yuta},
  booktitle={2014 IEEE International Conference on Robotics and Automation (ICRA)},
  pages={4620--4626},
  year={2014}
}

@book{hairer1993solving,
  author    = {E. Hairer and S. P. N{\o}rsett and G. Wanner},
  title     = {Solving Ordinary Differential Equations {I}: Nonstiff Problems},
  edition   = {2nd},
  publisher = {Springer},
  year      = {1993}
}

\end{document}